\documentclass{emulateapj}

\shorttitle{Variability in the HUDF}
\shortauthors{Cohen et al.}

\begin{document}

\title{Clues to AGN Growth from Optically Variable Objects\\
in the Hubble Ultra Deep Field}

\author{S. H. Cohen, R. E. Ryan Jr., A. N. Straughn, N. P. Hathi,
R. A. Windhorst}
\affil{Department of Physics and Astronomy, Arizona State University, Tempe, AZ
85287-1504}

\email{seth.cohen@asu.edu}

\and 

\author{A. M. Koekemoer, N. Pirzkal, C. Xu, B. Mobasher, S. Malhotra, L.-G.  Strolger,
J. E. Rhoads}
\affil{Space Telescope Science Institute, Baltimore, MD 21218}

\begin{abstract}

We present a photometric search for objects with point-source components that
are optically variable on timescales of weeks--months in the Hubble Ultra Deep
Field (HUDF) to $i'_{AB}\!=\!28.0$ mag. The data are split into four sub-stacks
of approximately equal exposure times. Objects exhibiting the signature of
optical variability are selected by studying
the photometric error distribution between the four different epochs, and
selecting 622 candidates as $3.0\sigma$ outliers from the original catalog of
4644 objects. Of these, 45 are visually confirmed as free of
contamination from close neighbors or various types of image defects. Four lie
within the positional error boxes of Chandra X-ray sources, and two of these
are spectroscopically confirmed AGN. The photometric redshift distribution of
the selected variable sample is compared to that of field galaxies, and we
find that a constant fraction of $\sim\!1\%$ of all field objects show
variability over the range of $0.1\!\lesssim\! z\!\lesssim\!4.5$. Combined
with other recent HUDF results, as well as those of recent state-of-the-art
numerical simulations, we discuss a potential link between the hierarchical
merging of galaxies and the growth of AGN.
\end{abstract}


\section{Introduction}

The Hubble Ultra Deep Field \citep[HUDF;][]{beck05} is the deepest optical image
of a slice of the Universe ever observed. As such, it allows for a variety of
different investigations into astrophysical objects and processes. The HUDF
observations consist of 400 orbits with the Hubble Space Telescope (HST) over
a period of four months in four optical bands ($BVi'z'$) with the Advanced
Camera for Surveys (ACS), and are supplemented in the $JH$ bands with an HST
Legacy Program using the Near-Infrared Camera and Multi-Object Spectrograph
\citep[NICMOS;][]{bouwens04, rtleg}. Given the unprecedented quality of these
data, the list of supporting data from both space and the ground is constantly
growing.

Since the ACS data was observed over a period of four months, it provides a
unique opportunity to search for variability in all types of objects to very
faint flux levels (perhaps even for $AB\gtrsim30$~mag), such as faint stars, 
distant supernovae (SNe), and weak active galactic nuclei (AGN). In this paper,
we search for weak AGN variability in the $i'$-band (F775W), because, in this
filter, the HUDF images are deepest, and have the best temporal spacing over
the four months for the variability search. It should be noted that at
higher redshifts, the ACS filters sample further into the rest-frame
ultraviolet. This is advantageous, because AGNs are known to show more
variability in the UV \citep[e.g.,][]{palt94}. In the original Hubble Deep
Field North (HDF--N) and the Groth Strip Survey, \citet{sara1} and
\citet{sara2} performed a similar search to AB=27.0 mag over longer time
baselines (5--7 years). They searched for variability in the nuclei of the
galaxies using small apertures, which is a different approach than in the
present search. 

From the WMAP polarization results \citep{kogut}, population III stars likely
existed at $z\!\simeq\!20$. These massive stars ($\gtrsim\!250$~$M_{\sun}$)
are expected to produce a large population of massive black holes
($M_{bh}\!\gtrsim\!150$~$M_{\sun}$, \citet{madrees}). Since there is now good
dynamical evidence for the existence of supermassive
($M_{bh}\simeq10^{6}\!-\!10^{9}$~$M_{\sun}$) black holes (SMBH) in the centers
of galaxies at $z\!\simeq\!0$ \citep{kor95,mag98,kor01}, it is important to
understand if there is any relationship between the formation of the SMBHs
observed at $z\!\simeq\!0$ and the lower mass BHs at z$\simeq$20. A
comprehensive review of SMBHs is given by \citet{ff04}. An important
question to address is how these SMBHs, as seen nearby, have grown over
the course of cosmic time. One suggestion is that they ``grow'' through the
mergers of galaxies that themselves contain less massive BHs, so the byproduct
is a larger single galaxy with--eventually--a more massive BH in its center.
The growth of the BH may then be observed via its AGN activity. If this
scenario is true, then perhaps there exists an observable link between galaxy
mergers and increased AGN activity \citep{silkrees}. Therefore, studying
this possible link as a function of redshift could give insight into the
growth of SMBHs and its potential relation to the process of galaxy assembly.

In \S 2, we present the HUDF observations and summarize the essential elements
of its data reduction, and in \S 3 we present the variable candidate selection.
In \S 4 we present the photometric redshift distribution of the variable objects
together with that of the HUDF field galaxies, in \S 5 we present the results,
and in \S 6 we discuss our results in terms of galaxy assembly and AGN growth.

\section{Observations}

All data used here are from the Hubble Ultra Deep Field \citep[HUDF;][]{beck05}.
The individual cosmic-ray (CR) clipped images and weight maps were used with
{\it multidrizzle} \citep{driz} to create eight sub-stacks of approximately
equal exposure times. These used the same cosmic-ray maps and weight maps
employed to create the full-depth HUDF mosaics. All HUDF images were {\it
drizzled} onto the {\it same} output pixel scale ($0\farcs030$ per pixel) and
WCS frame as the original HUDF. Given the time-spacing of the exposures and the
desire to extend the study to the faintest possible flux-levels, the images and
weight maps were combined in groups of two to create four epochs of observation
for the variability study on 0.4--3.5 months timescales. Exposure--time weighted
averages were created for all images, and simple addition was used to combine
the weight maps \citep{beck05}. The exposure times and median observation dates
are listed in Table~\ref{tab1}. The four epochs chosen here have exposure times
as close to each other as possible, so that the flux error distributions will be
as much as possible symmetric, and therefore more easily modeled. As seen in
Table~\ref{tab1}, this is done at the expense of not having the endpoints of 
the four epochs well-spaced in time. In order to be optimized for variability
studies, future observations of this kind should take
into account the need for both equal depth exposures and well separated
observation dates, although this would further complicate the already difficult
task of scheduling observations such as the HUDF. All magnitudes are on the
$AB$-system using the zero-points
given in the HUDF public data release \citep{beck05}.

\begin{table}
\caption{Observations}
\label{tab1}
\begin{tabular*}{0.51\textwidth}
   {@{\extracolsep{\fill}}ccccccc}
\hline
\hline
\multicolumn{1}{c}{Epoch} & \multicolumn{1}{c}{Start} & \multicolumn{1}{c}{End} &\multicolumn{1}{c}{Exp. Time\tablenotemark{a}} & \multicolumn{1}{c}{\# of} &\multicolumn{1}{c}{RJD\tablenotemark{b}} & \multicolumn{1}{c}{Days Since}\\
 \multicolumn{1}{c}{No.} & \multicolumn{1}{c}{Date} & \multicolumn{1}{c}{Date} & \multicolumn{1}{c}{(Seconds)} & \multicolumn{1}{c}{Exp.} &\multicolumn{1}{c}{(Days)}& \multicolumn{1}{c}{Epoch 1} \\
$ $&$ $&$ $&$ $&$ $&$ $&$ $\\
\hline
1 & 2003-09-24 & 2003-10-10 &92340 & 76 & 52914.2 & \nodata \\
2 & 2003-10-10 & 2003-10-29 &92340 & 76 & 52926.7 & 12.5 \\
3 & 2003-12-04 & 2003-12-18 &89940 & 76 & 52985.9 & 71.7\\
4 & 2003-12-22 & 2004-01-14 &72490 & 60 & 53005.7 & 91.5\\
\hline
\end{tabular*}
\tablenotetext{a}{There are two exposures per HST orbit}
\tablenotetext{b}{RJD=median Revised Julian Date$-2.4\!\times\!10^6$ days}
\end{table}

\subsection{Catalog Generation and Photometry \label{catgen}}

Variability was searched for by comparing the photometric catalogs from the 
various epochs to each other. Catalogs were generated using {\it SExtractor 
Version 2.2.2} \citep{ba96} with a $1.0\sigma$ detection threshold, and
requiring a minimum of 15 connected pixels (i.e. aproximately the PSF area)
at this limit above sky. Since we are
searching for any signs of variability, we chose to use a liberal amount of
de-blending ($DEBLEND\_MINCONT=5\!\times\!10^{-6}$). This allows for pieces of
merging galaxies to be measured separately to enhance the chances of finding
variable events in point-source components. {\it SExtractor} was run in dual
image mode using the full-depth HUDF as the detection image, and utilizing the
corresponding weight-maps to minimize the number of false detections due to
edges and other image defects that are reflected in these weight-maps. This
results in catalogs with 27819 measured objects, which still contains many
over-deblended objects or edge-effects. Since the HUDF was observed at four
different position angles \citep{beck05}, only the 15205 objects observed in
all four epochs were considered. The result is a catalog of 12514 objects,
which is 90\% complete to $i'\!\lesssim\!30.5$~mag. Since
we can only measure variability from the individual epoch images that are
one-fourth the full HUDF in length, the variable candidate sample is restricted
to the 4644 objects with $i'\!<\!28.0$~mag. 

The HUDF is the deepest optical image ever observed, and will possibly 
remain the deepest until the James Webb Space Telescope (JWST) is launched in
2011. To explore the limits of the HUDF depth, a few words about the point
spread function (PSF) of the ACS images are needed. A comprehensive study of
HST/ACS PSF-issues can be found in \citet{hey04}, so here we only highlight the
aspects relevant to our study. First, the ACS camera is not located on the
optical axis of HST, and therefore the HUDF field is rectified by applying
geometric distortion correction polynomials. Secondly, the ACS/WFC PSF is
known to vary with location on the CCD detectors, and with the time of
observation due to ``breathing'' of the HST Optical Telescope Assembly (OTA).
Since the HUDF was observed at four different roll angles over four months,
these PSF effects can easily be seen by inspecting the locations of bright
objects in an image created by dividing two images taken at different roll
angles. Owing to these PSF issues and the significantly complex ACS image
registration, this ``ratio-image'' will easily show the cores of all bright
objects with significant positive or negative flux excursions, regardless of
whether or not they are truly variable. For this reason, we {\it cannot} use
small PSF-sized apertures to search for nuclear variability, as was done by
\citet{sara1}. \citet{sara1} {\it could} use the small--aperture method,
because for the much larger WFPC2 pixel-size and the {\it on-axis} location of
the WFPC2 camera, the geometrical distortion correction and registration effects
are much smaller. Instead, we chose to use total magnitudes of the highly
deblended objects. Even though our total flux apertures may encompass the whole
galaxy, the variability necessarily must come from a small region (less than
the $0\farcs084$ PSF), due to the finite light-travel time across the region of
variability. 
\section{Candidate Selection} \label{select}

The catalogs of the four epochs were all compared to each other resulting in six
sets of diagrams similar to the one shown in Fig.~\ref{tree}. These figures
show the change in measured magnitudes in the SExtractor matched apertures as
a function of full-depth HUDF flux. In order to determine which objects were
true outliers (i.e., variable candidates), the expected error distribution
for each set was computed as follows. For each measurement in a given epoch,
we compute the total flux error for that $i^{th}$ flux measurement:

\begin{figure}
\epsscale{1.2}
\plotone{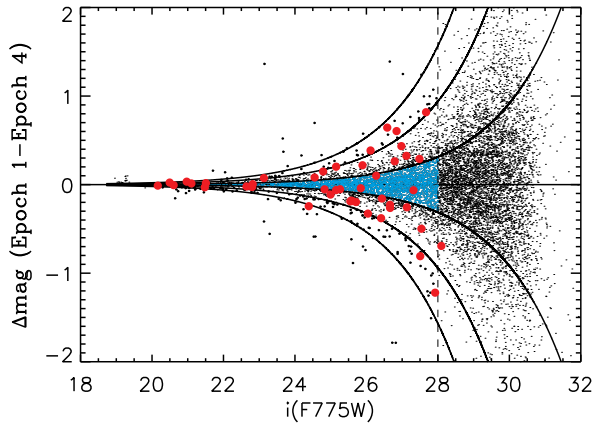}
\caption{Magnitude difference between two HUDF epochs of all
objects versus their total $i'$-band flux in AB-mag from matched total
apertures. The $\pm 1 \sigma,\pm 3\sigma, \pm 5\sigma$ lines are shown. The
black circles show the 222 variable candidates chosen to be $| \Delta mag |
\gtrsim\!3\sigma$ outliers between any one of these these two epochs and $20\!
\lesssim\! i' \!\lesssim\! 28.0$~mag. Blue points show the $| \Delta mag |
\leq\!\pm\!1\sigma$ points used to normalize the error distribution, so that
1.0$\sigma$ reflects as much as possible the true Gaussian 1-$\sigma$ line. 
Large red points show the ``best'' 45 candidates that were chosen from all six
possible epoch combinations, many of which were seen at $\gtrsim\!3.0\sigma$
in 2 or more epoch combinations.  \label{tree}}
\end{figure}

\begin{equation}\label{err1}
\sigma_i^{tot}=\sqrt{\frac{\sigma_i^2 A(F_i)+F_i/G}{F_i^2}}
\end{equation}

\noindent where $\sigma_i$ is the RMS per pixel in the sky background, $A(F_i)$ is the
number of pixels belonging to object $i$ of a given flux (described in detail
below), $F_i$ is its measured flux in $e^{-}$ per second and $G$ is the gain in
units of seconds. This quantity is computed for each epoch, and for each of
the epoch-pairs they are combined in quadrature: 

\begin{equation}\label{errtot}
\Delta mag=\pm\sqrt{(1.0857N)^2\times\left((\sigma_i^{tot})^2+(\sigma_j^{tot})^2\right)}
\end{equation}

\noindent where $N$ is the number of $\sigma$ for which the error-model is
computed, and $i,j$ denote the measurements at a given flux-level in each
of the two epochs under consideration. Since each object at a given flux level
can have a different area $A$ (i.e., number of pixels), we need to assume a
general relation between flux and area in order to optimally model the true
error distribution using the above equations. This relation was determined
iteratively for each pair of observations, such that
68.3\% of the points lie within the boundaries of the upper and lower
$1.0\sigma$ lines. We started this process by fitting the relation between the
{\it SExtractor} magnitude and {\it ISOAREA\_IMAGE} parameters as a first
guess. It is then assumed that flux is proportional to area, and we solve
iteratively for the proportionality constant to get $\pm$1.0$\sigma$ lines that
maximally represent a Gaussian error distribution (Fig.~\ref{tree}). In order
to demostrate the Gaussian nature of this error distribution at all flux
levels, the $\Delta mag$ data are divided by the 1.0--$\sigma$ model line,
and histograms were computed for the resulting $\Delta mag$ data at various
flux-levels
(Fig.~\ref{gauss}). These histograms are remarkably well fit by Gaussians
of $\sigma\!\simeq\!1$. The HUDF noise distribution is not perfectly Gaussian,
but with 288 independent exposures in the $i'$-band, the error distribution
is as close to Gaussian as seen in any astronomical CCD application.

\begin{figure}
\epsscale{1.2}
\plotone{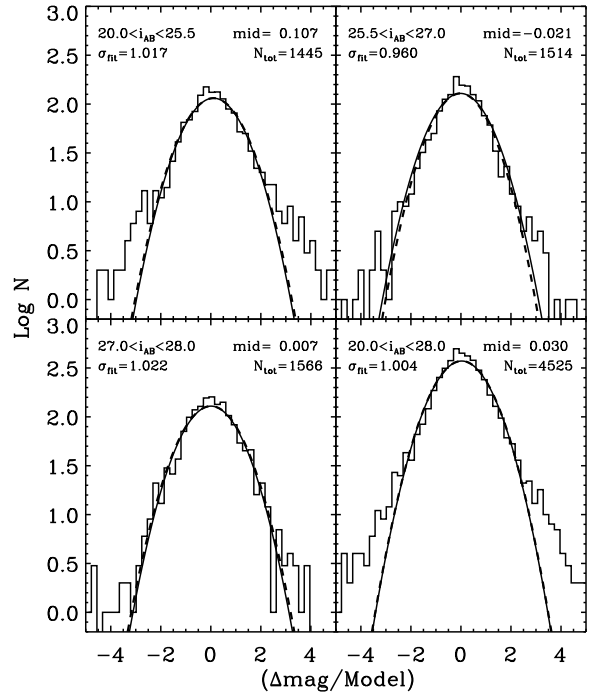}
\figcaption[f2.eps]{Gaussian nature of the flux error distribution at all flux
levels. The $\Delta mag$ data from Fig.~\ref{tree} are divided by the best--fit
model 1.0-$\sigma$ lines. Histograms are computed from the resulting data for
the indicated magnitude ranges. All distributions are well fit by Gaussians
(parabolas in log space) with $\sigma\!\simeq\!1.0$ as indicated in the 
individual panels. The almost indistinguishable dashed and solid lines
are for the best--fit and $\sigma\!\equiv\!1$ Gaussians, respectively.
\label{gauss}}
\end{figure}

Once this $1\sigma$ line is determined, we set $N\!=\!3.0$ in
Eqn.~\ref{errtot}, and find all the objects that are at least $3.0\sigma$
outliers. In Fig.~\ref{tree}, we show the $\pm$$1\sigma$, $\pm$$3\sigma$,
and $\pm$$5\sigma$ lines, along with the actual data. For this particular pair
of catalogs, there were initially 222 objects which
varied in flux by more than $3.0\sigma$. The choice of $3.0\sigma$ implies that
we can expect 0.27\% random contaminants. Among 4644 objects to $i'\!=\!28.0$~mag,
this is about 13 interloping objects that are potentially classified as
variable, simply because of the chosen $3.0\sigma$ cut-off. 

\section{Photometric Redshifts}

Photometric redshifts were computed in two ways. In the first method, magnitudes
are computed in a fixed aperture with a one arcsecond radius, and objects are
selected from the ACS $z'$-band. In addition to the HUDF ACS $BVi'z'$ images,
the available NICMOS
$JH$ \citep{rtleg} and VLT ISAAC $K$ band images are used where available. The
$z'$-band selection allows for $z\!\gtrsim\!5.5$ galaxies to be included in the
list, but the $z'$-band is not as deep as the $i'$-band or $V$-band images in
the HUDF. Therefore, the primary object definition catalog was made in the
$i'$-band, which introduces a bias against $z\!\gtrsim\!5.5$ objects (see Yan
\& Windhorst 2004). The use of the large apertures allows for the ground-based
seeing $K$-band fluxes to be included for more precise photo-z estimates. One
possible problem is that most faint galaxies are significantly smaller than
these apertures, such that problems may arise for objects in crowded regions.

In an attempt to address these issues, we also tried using magnitudes measured
within apertures defined in the $i'$-band, using the same apertures in which the
variability was searched for ({\it SExtractor} parameter {\it MAG\_AUTO}).
The $i'$-band selection limits us to objects with $z\!\lesssim\!5.5$.
This photometry is only applied to the ACS $BVi'z'$ and NICMOS $JH$ data,
which have the necessary resolution to accurately measure fluxes on
sub-arcsecond scales. The VLT $K$-band data are not used here, because they are
limited by ground-based seeing ($\lesssim$ 1'' FWHM). The
disadvantage of not using the $K$-band is the lower redshift accuracy, but the
advantage is that the flux is measured from the {\it same} object component in
all filters, so that crowding is less of an issue for this method. The fluxes
and errors measured in this way are then input into the photometric redshift
code {\it hyper-z} \citep{hyper}, using a suite of both empirical \citep{cww}
and evolutionary spectral synthesis \citep[GISSEL98 update to][]{bc93}
templates.

While these methods each have their own advantages and disadvantages, they
produced the same important results discussed below. The difference between
the methods are minor in the photometric redshift distribution produced, and
since we discuss ratios of photometric redshift distributions in what follows,
these differences are not relevant for the main argument below. The second
photometric redshift determination method was adopted for all figures
shown here.

\section{Results}

\subsection{Number of Variable Objects}

\begin{figure}
\epsscale{1.2}
\plotone{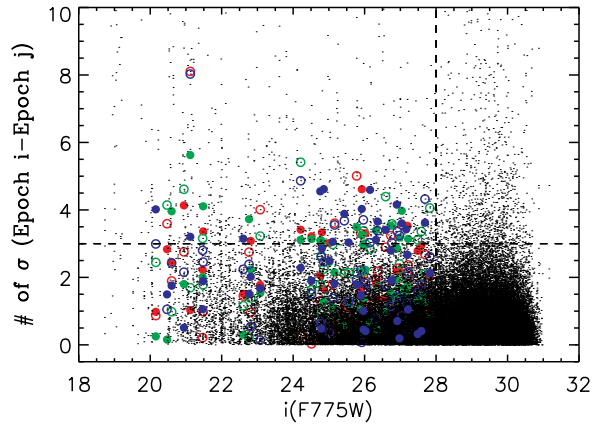}
\figcaption[f3.eps]{Number of $\sigma$ that each object varies for all
six possible combinations of the four different epochs. The ``best'' sample of
45 variable candidates is shown with colored symbols. These objects were
chosen to be $| \Delta mag | \!\gtrsim3.0 \sigma$ outliers, have $20\!
\lesssim\! i' \!\lesssim\! 28.0$~mag, and to be unaffected of any local image or
weight map structures. Each object appears six times in this plot, and most
candidates were seen at $\gtrsim\!3\sigma$ in at least two epoch pairs.
Significant outliers not plotted in color were almost exclusively due to
over-splitting or deblending issues with large objects (mostly occurring in
bright, large spiral galaxies), where the enhanced uncertainties in the local
(object+sky)-subtraction introduced larger--than--Gaussian flux errors. 
\label{sigplot}}
\end{figure}

We find unique 622 out of 4644 variable candidates with $i'\!\leq\!28.0$~mag
from the six possible 2--epoch combinations. Of these, 66 are rejected as
having $i'\!>\!28.0$~mag in the final HUDF stack, leaving only 556 candidates
to our magnitude limit. An object where just a single point in the light-curve
is deviant would appear as an outlier in 3 out of 6 epoch combinations. This
occurs in 25\% of the candidates. Another 25\% stand out in 2 out of 6 pairs,
and 40\% stand out in 1 out of 6 pairs (usually indicative of a global rise
or decline as a function of time in the light-curve). Of these 556 initial
candidates, we find that only 45 out of 4644 show a clear sign of a compact
region indicative of a point source, and are devoid of image defects or object
splitting issues. These object crowding or splitting issues arise due to the 
extreme deblending threshold (see \S~\ref{catgen}), which causes unreliable
detections and photometric measurements of faint objects in the wings of
bright objects. Therefore, we have 45 of 4644 objects that show the signature
of AGN variability. The 13 interlopers discussed in \S~\ref{select} should,
in the absence of other information, be evenly distributed amongst the 622
initial candidates, and therefore we expect on the order of 1 out of 45 of our
candidates to be a random contaminant. In order to simultaneously show the
data from all six possible combinations of the data, the results are plotted in
Fig.~\ref{sigplot}, showing the number of $\sigma$ by which each object
varied in each epoch pair.  The colored symbols are for the 45 ``best''
candidates. 

\begin{figure*}
\centerline{\epsfxsize=\hsize{\epsfbox{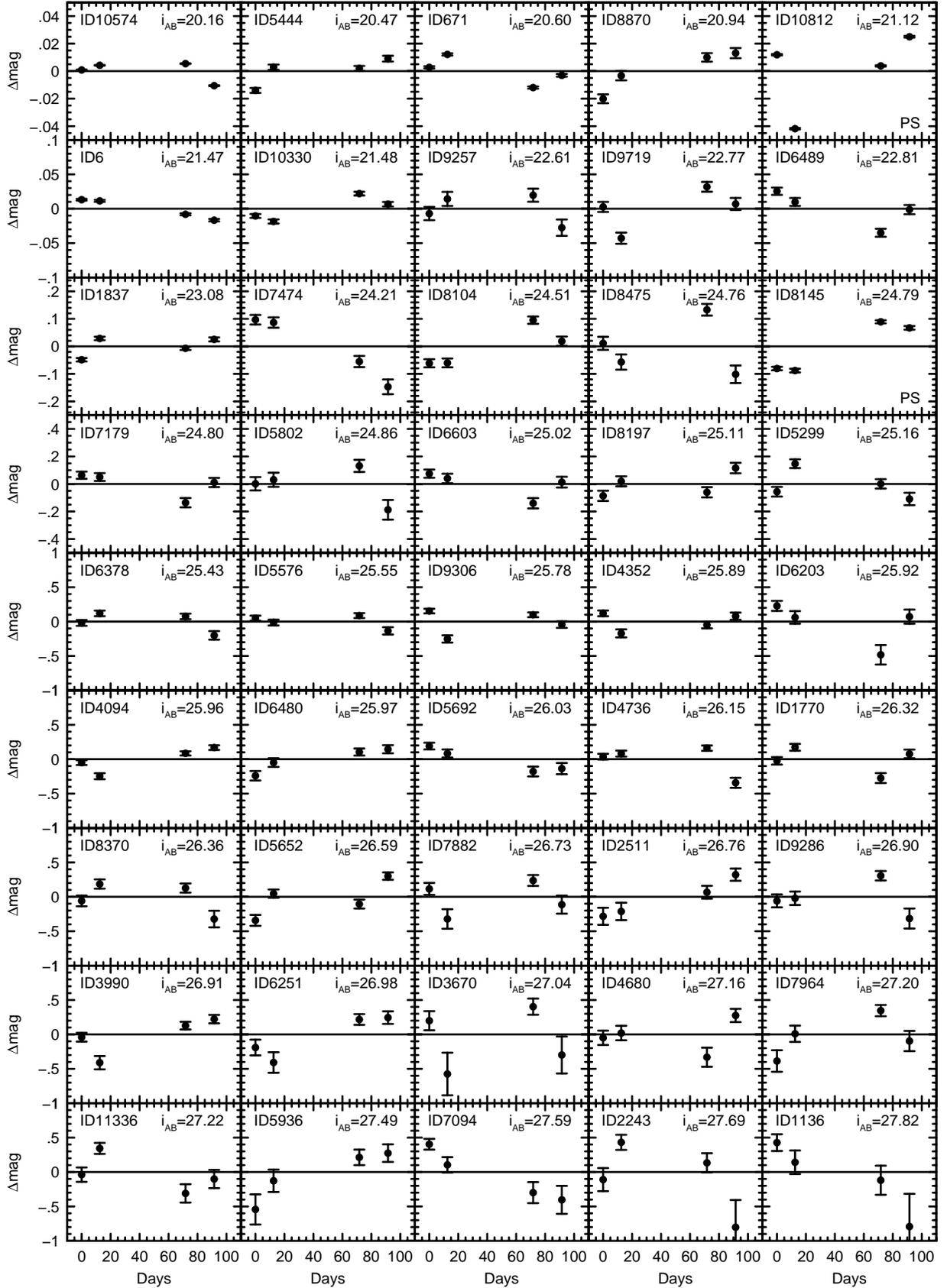}}}
\caption{Light curves of the 45 best candidates with signs of
optical AGN variability. The vertical axes are the change in measured magnitudes
plotted in the sense of the average minus individual measurements. The
horizontal axis is the number of days since the first measurement. Each panel
shows the object ID number in the upper left, and the $i'_{AB}$ magnitude
in the upper right. Plots are arranged in order of decreasing flux, and the
combined total flux
error bars are from {\it SExtractor.} The two point sources discussed
in \S~\ref{otherwave} are indicated by ``PS.'' \label{curves}}
\end{figure*}

Another 57 objects were found that are ``potentially'' variable candidates.
These are relatively isolated objects with reliable photometry, but show no
clear sign of a point source. Since variability has to come from a point-source
due to light-travel time considerations, we will ignore these 57 objects for
now, but in \S 6 we will discuss the incompleteness resulting from the finite
variability timescales sampled, and from the fraction of non-variable or
dust-obscured AGN that have likely been missed altogether. The four-epoch
light-curves for the 45 best candidates are shown in Fig.~\ref{curves}.
These light-curve data are tabulated in Table~\ref{tab2}, which also
specifies the number of epoch pairs where each object was a 3.0$\sigma$ outlier.
For the 45 best candidates, 49\% were discovered from a single pair, and 43\%
in two pairs. Only 5\% (2 objects) were found in 3 pairs, which is indicative of
a single deviant point in the 4 point light-curve. 

\begin{figure}
\epsscale{1.2}
\plotone{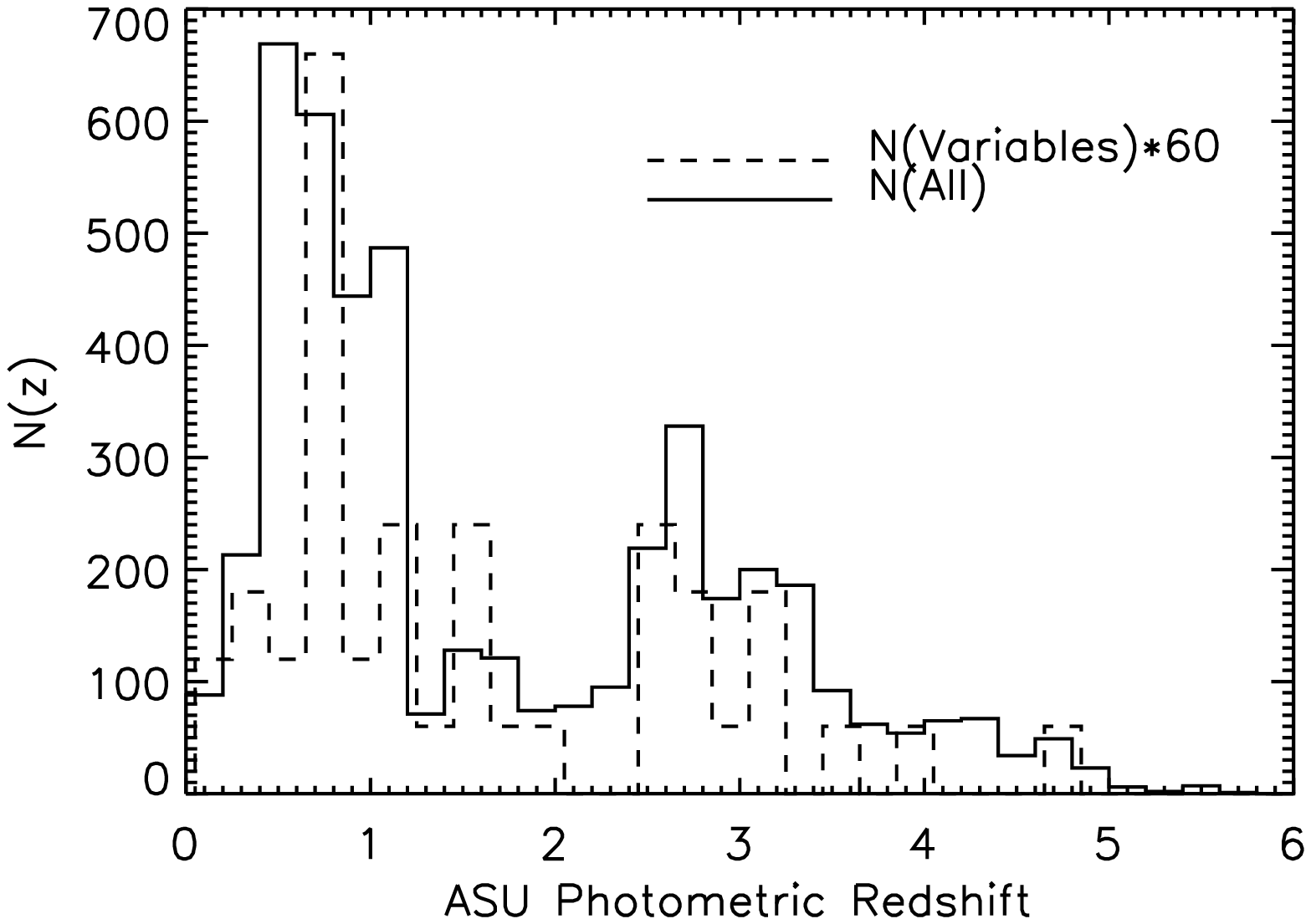}
\figcaption[f5.eps]{Photometric redshift distribution of all HUDF
field galaxies (solid line) and for the ``best'' variable candidates (dashed
line) multiplied by 60$\times$ for best comparison. Photo-z's computed using
$hyperZ$ \citep{hyper} and $BVi'z'JH$ HST data for all galaxies with
$i'\!\lesssim\!28.0$~mag. The redshift distribution of the variable objects
follows that of field galaxies in general. 
\label{zdist}}
\end{figure}

\begin{figure}
\epsscale{1.2}
\plotone{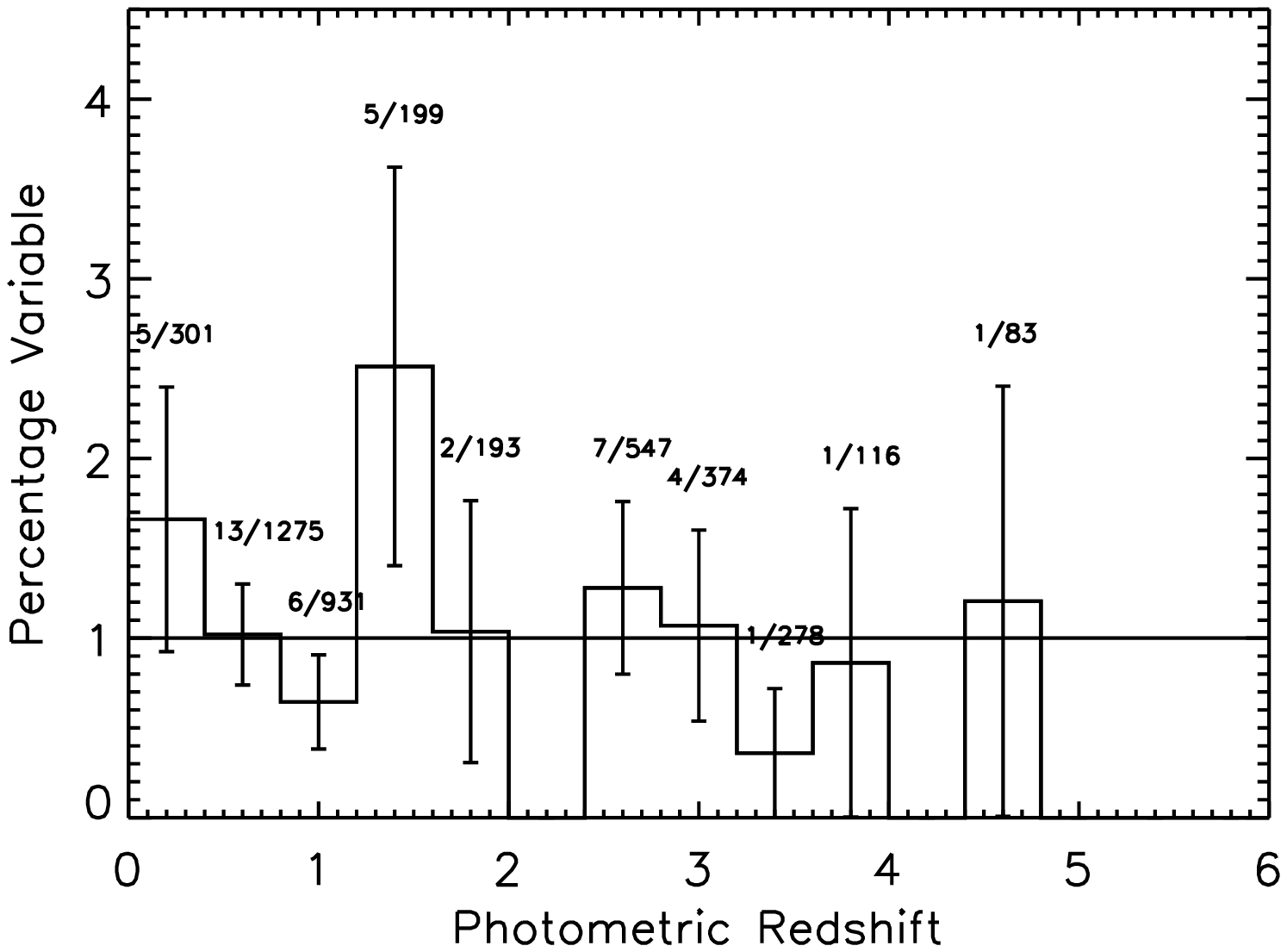}
\figcaption[f6.eps]{Percentage of HUDF objects to 
$i' \! \lesssim\!28.0$ AB-mag showing variable point sources
as a function of photo-z. Within the statistical uncertainties,
about 1\% of all HUDF galaxies show point source variability over the whole
redshift range surveyed ($0\!\lesssim\!z\!\lesssim\!5$). \label{zratio}}
\end{figure}

In Fig.~\ref{zdist}, we show the photometric redshift distribution for all
objects with $i'\!<\!28.0$~mag along with that of our best 45 candidates. It is 
clear that the distribution follows that of the field galaxies, i.e., there is
no redshift where faint object variability was most prevalent. To make this
more clear, we plot in Fig.~\ref{zratio} the ratio of the $N(z)$ for
variables to that of field galaxies, which shows that this fraction is
roughly constant at approximately 1\% over all redshifts probed in this study.
This variability fraction is similar to the 2\% found by \citet{sara2} in a
search for nuclear variability in the HDF--N and the Groth Strip survey. 

Interestingly, \citet{tadpole} show, in a companion study, that the redshift
distribution for ``tadpole'' galaxies also traces that of the HUDF field
galaxies. They argue that these tadpole galaxies are dynamically unrelaxed
systems, and therefore in the early stages of merging. The question then
arises if tadpole galaxies and objects with point-sources that show signs of AGN
variability are drawn from the same population. This question is further
addressed in \S~\ref{discuss}.

\subsection{Other Tests of the Reliability of the Variable Sample}

\subsubsection{Difference Images}

We attempted to also detect the point-source variability using ``difference''
or ``ratio'' images between any of the combinations between the two epochs. To
do this, we smoothed the image by a $5\!\times\!5$ and $7\!\times\!7$ box. For each
object, a ``variance map'' is computed over the four epochs as follows. For
each pixel, we compute the $n\!\times\! n$ box-average for both the image and the
weight map. Then we compute the weighted standard deviation over the four
epochs for {\it each} pixel in the box-averaged images. The $n\!\times\! n$ box-size was
chosen to smooth out most of the PSF breathing and registration issues. Even
so, this variance image clearly had peaks in the centers of all bright
objects. However, the variable candidates also tended to stand out more than
their non-variable neighbors. This method worked best for the brighter objects,
clearly verifying 13/18 candidates with $i'_{AB}\!\lesssim\!25$~mag, but
only finding $\sim$50\% of the full $i'_{AB}\!\lesssim\!28$~mag sample. While
not an optimal method of detection, this
ratio-method provided a good way of confirming some of the objects whose total flux
changed significantly over the four month period. However, the total flux
method remains our primary variable candidate detection method, since it is
most robust against ACS PSF effects.

\subsubsection{Variability Between Two Deeper Epochs}

Given the observing dates listed in Table~\ref{tab1}, a natural test is to
combine the first and second epochs, and also the third and fourth, giving only
two independent HUDF epochs but more widely spaced in time, and with a somewhat
higher signal-to-noise ratio for the variable candidate detection. However, we
expect fewer variable candidates this way, because there are fewer epochs to
compare, and because some of the short-term variability signature is
necessarily smoothed over. From this
2--epoch test, 242 candidates are chosen from the error distribution. Next this
list is compared to the list of the 45 best candidates from the 4-epoch search,
and 12 objects were found in both searches. If the list of the best and possible
candidates are combined, there are 102 candidates, and 30 of these are
recovered in the two epoch test. This test gives further confidence that at
least 30\% of our top candidates are truly variable objects, although it does
not exclude any of the other ones, since those were found when the shorter 
time-baselines were included. 

A sample of local AGN light-curves can be used in order to assess the
expected completeness in temporally condensing our HUDF data from four to
two epochs. The best publicly available data is that from the extensive
International AGN Watch\footnote{Tabulated data from the International AGN
Watch can be found at the URL
\url{http://www.astronomy.ohio-state.edu/$\sim$agnwatch/}}
\citep[hereafter IAGNW;][and references therein]{pete02}. We used their five
IAGNW Seyferts (all with $z\!<\!0.05$), which have $B$-band light-curves which
best match our observed HUDF $i'$-band data, since the median redshift of our
candidates is $z_{med}\!\simeq\!1.5$. Since the rest-frame time sampling depends
on the redshifts  of the HUDF objects, we resampled the IAGNW light-curves for a
range of redshifts between $0.5\!<\!z\!<\!5$. The IAGNW data allowed for 30--40
independent 4--epoch light-curves to be created for a given redshift. These
light-curves were then scaled to match the observed spread in $\Delta mag$ as
a function of magnitude for our 45 HUDF candidates. A Monte Carlo test was then
run on mock catalogs that matched the same magnitude distribution as our 45
HUDF candidates. On average half of the $\sim\!40$
light-curves are found when our 4--epoch variability algorithm (using same
$3\sigma$ selection curves) is applied. The 2--epoch algorithm is then applied to
those mock candidates and the number that remain variable candidates turns out
to be a decreasing function of assumed redshift. We recover 60--70\% for
$z\!\lesssim\!1$ and 30--40\% for $z\!\gtrsim\!4$. However, this test relies on
using 20 templates to simulate 45 objects. A visual inspection of the
light-curves in Fig.~\ref{curves} shows that ideally a larger number of local
templates is needed for a fair test. Nonetheless, this limited test that is
possible with the available IAGNW data -- when taken at face value -- may be an
indication that as many as one-third of our
45 candidates are potentially false positives, although it is equally likely that
the small number of local test templates are not truly representative of the 
distant HUDF 
variability sample. Hence, we believe that the reliability of our faint HUDF
variability sample is likely larger than 67\%, but we will not be able to 
definitively say for sure without more local light-curve templates that best 
match the high-z variable objects.

\subsubsection{Other Wavelengths} \label{otherwave}

Since the UDF is in the Chandra Deep Field--South \citep[CDFS;][]{rosati02},
there exists deep X-ray data in this field, and these data are ideal for
detecting AGN. Within the UDF, there are 16 Chandra sources 
(Koekemoer et al. 2005, in preparation), and we detect four of these as
variable in the optical. One of these is a mid-type spiral with
$i'\!=\!21.24$~mag, that is a member of a small group of apparently interacting
galaxies. Interestingly, this object's total flux varied by less than 1\%, but
since it was one of the brightest objects in the sample, its variability was
clearly detected at $\gtrsim\!3.0 \sigma$ significance. Two of the others
are optical point sources, showing little or no visible host galaxy,
with $i'\!=\!21.12$~mag and $i'\!=\!24.79$~mag. They both
have measured spectroscopic AGN emission-line redshifts from the GRism ACS
Program for Extragalactic Science survey \citep[GRAPES;][]{grapes} at
$z\!\simeq\!3$. Our $BVi'z'JH$ photometric redshifts put them at $z\!=\!2.6$
and $z\!=\!2.9$ respectively, which is rather good given that these few
brighter AGN are the ones that are most dominated by a non-stellar power-law
SED. The fourth candidate is barely detected in any of the ACS bands, and
would have been eliminated as a spurious candidate had it not been coincident
with the Chandra source, and is clearly a viable object in the VLT $K$-band
image. The detection of 25\% of the Chandra sources as optically variable in
the HUDF data shows that the method employed here is a reliable way of finding
the AGN that are not heavily obscured. \citet{pao04} find that $>\!90\%$ of 
the studied CDFS X-ray sources with adequate photon statistics show X-ray 
variability, and a future comparison of the X-ray and optical variability 
amplitudes could provide insight into the physical mechanism that is 
responsible for this  variability.

\subsection{Possible Sources of Incompleteness in the Variable AGN Sample} 

In summary, we found 45 plausibly variable objects with some caveats, and this
number may be as high as 102 if the ``possible'' variable candidates are 
included. Hence, the true variability fraction on timescales of a few months
(rest-frame timescale few weeks to a month) is no more than about 1--2\% of all
HUDF field galaxies. Except for variable stars such as SNe and novae, this
variability is most likely due to an AGN given the timescales and 
distances involved.
\citet{sr05} found only one moderate redshift SNe in the HUDF, and thus
SNe cannot be a significant source of contamination of the present sample.

Two other possible source of incompleteness in the variability study must be
addressed first. Non-variable AGN, or AGN that only vary on timescales much
longer than 4 months, or optically obscured AGN would not have been detected
with the current UV--optical variability method. Sarajedini et al. (2003a) had
epochs 5--7 years apart, and found 2\% of the objects to be variable. It is
thus possible that the currently sampled times-scale shorter than 4 months
missed about a factor 2 of all AGN that are only variable on longer
time-scales. 

A factor of three of the brighter AGN may have been missed, since their
UV--optical flux may be obscured by a dust-torus (Barthel 1989). In the AGN
unification picture, AGN cones are two-sided and their axes are randomly
distributed in the sky, so that an average cone opening angle of $\omega$
implies that a fraction 1--sin($\omega$) of all AGN will point in our direction.
If $\omega\!\simeq\!45^\circ$ (e.g., Barthel 1989), then every optically detected AGN
(QSO) represents 3--4 other bulge-dominated galaxies, whose AGN reflection cone
didn't shine in our direction. Hence, their AGN may remain obscured by the
dust-torus. Such objects would be visible to Chandra in the X-rays or to Spitzer
at mid-IR wavelengths, although the available Chandra and Spitzer data are not
nearly deep enough to detect all HUDF objects to $AB\!=\!28$~mag. (Reaching
these depths is prohibitive in Chandra and Spitzer integration
times, and requires the next generation of X-ray and IR telescopes, such as
Generation-X and JWST). At brighter flux limits, Spitzer did indeed recently
find a significant fraction of dust-obscured AGN not seen in UV-optical surveys
\citep{urry04}. In the AGN unification picture, the incompleteness in
UV--optically selected samples due to the
dust-obscuring torus would be as large as a factor of 3--4 \citep{treister04}.

\section{Discussion and Conclusions} \label{discuss}

\subsection{Fraction of Variable AGN found}

Interestingly, \citet{tadpole} show, in a companion study, that the redshift
distribution for ``tadpole'' galaxies also traces that of the HUDF field
galaxies for $0.1\!\lesssim\! z\! \lesssim\! 4.5$. They argue that these tadpole
galaxies are dynamically unrelaxed
systems, and therefore in the early stages of merging. The question then
arises if tadpole galaxies and objects with point-sources that show signs of AGN
variability are drawn from the same population. At any given redshift,
\citet{tadpole} find that about 6\% of all HUDF galaxies appear as tadpoles, 
and they conclude that tadpole galaxies are good tracers of the process
of galaxy assembly.

Together with the factor of $\gtrsim\!2$ incompleteness in the HUDF variability
sample due to the limited time-baseline sampled thus far, the actual fraction
of weak AGN present in these dynamically young galaxies may be a factor of
$\gtrsim$6--8$\times$
larger than the 1\% variable AGN fraction found through variability in the
HUDF. Hence, perhaps as many as $\gtrsim$6--8\% of all field galaxies may host
weak AGN, only $\sim\!1\%$ of which we found here, and another $\gtrsim$~1\%  could
have been found if longer time-baseline had been available. The other factor of
3--4$\times$ of AGN are likely
missing because they are optically obscured, requiring the next generation of
X-ray and IR telescopes. 

\subsection{AGN feeding timescales compared to galaxy merger timescales} 

We now consider if the current variability dataset can constrain the merging
rate of SMBHs, assuming that SMBHs formed during hierarchical mergers of
galaxies containing less massive SMBHs, and if signatures of galaxy mergers
could be related to variable AGNs.  Put another way,
can signatures of galaxy merging be related to our variable AGN candidates? 

A closer inspection of our data reveals that only 1 or perhaps 2 of the variable
candidates resemble the tadpole galaxies of \citet{tadpole}. Recent
state-of-the-art hierarchical models have suggested that the AGN-phase {\it
only} occurs in the later stages of a galaxy merger, and well {\it after} it
appears in the tadpole phase \citep{springel05, dimatteo05}. These models also
predict that the AGN will likely only be visible {\it well after} ($\gtrsim$1
Gyr) the merger induced star-formation has died down, implying that the
fraction of dynamically young tadpoles that are expected to already show active
AGN properties is relatively small. The small overlap between the two
populations that we observe does thus not exclude the possibility that faint
object variability is tracing the growth of SMBHs. However, this SMBH growth
can only be constrained indirectly from the variability data, and a more 
detailed discussion of the connection between mergers and AGN activity is 
given in \citet{tadpole} and references therein. 

Given the importance of understanding the growth and origin of
SMBHs, several lessons can be learned from this work in order to better design
future studies of this type. The time-spacing of the HUDF observations
--although as good as possible given scheduling constraints--was not ideal
for this type of study. It is critical to re-visit the HUDF with HST, with the
observations optimized for a faint-object variability study, including
covering time-scales of a few years. It is also essential to plan deeper
surveys at longer wavelengths with the JWST. The JWST photometric 
{\it and} PSF stability are crucial in this regard, as many of our HUDF 
objects show significant variability of less than a few percent in flux. Also, a limiting
factor in our results is the breathing of the PSF, along with image noise due
to correcting the geometric distortion of ACS and slight variations of the PSF
across the field. Therefore, JWST must have design specifications that are
capable of meeting these requirements in order to permit faint object
variability studies to be done.

\acknowledgments

We acknowledge the support from the NASA grants HST-GO-09793.*, awarded by
STScI, which is operated by AURA for NASA under contract NAS 5-26555. This work
was funded in part by NASA JWST grant NAG5-12460 (to RAW). This work
benefited from fruitful discussions with Drs. M. Corbin, H. Ferguson,
L. Hernquist, R. Jansen, V. Sarajedini, and many others. We also thank the
anonymous referee for helpful suggestions that improved the paper.

{\it Facility:} \facility{HST(ACS)}.

\clearpage

\begin{table}
\caption{Best 45 Variability Candidates}
\label{tab2}
\begin{tabular*}{0.98\textwidth} {@{\extracolsep{\fill}}rrrrrrrrrrl}
\hline
\hline
\multicolumn{1}{c}{ID} & \multicolumn{1}{c}{RA (J2000)} & \multicolumn{1}{c}{Dec (J2000)} & \multicolumn{1}{c}{$i_{AB}$\tablenotemark{a}} &
\multicolumn{1}{c}{$\langle i_{AB}\rangle$\tablenotemark{b}} & \multicolumn{1}{c}{$\Delta mag1$\tablenotemark{c}} & \multicolumn{1}{c}{$\Delta mag2$\tablenotemark{c}} &
\multicolumn{1}{c}{$\Delta mag3$\tablenotemark{c}} &\multicolumn{1}{c}{$\Delta mag4$\tablenotemark{c}} & \multicolumn{1}{c}{\#\tablenotemark{d}} &\multicolumn{1}{c}{Notes\tablenotemark{e}} \\
\multicolumn{1}{c}{} & \multicolumn{1}{c}{({\it h m s})} & \multicolumn{1}{c}{(\degr \, \arcmin \, \arcsec)} &
\multicolumn{1}{c}{(mag)} &\multicolumn{1}{c}{(mag)} &  \multicolumn{1}{c}{(mmag)} & \multicolumn{1}{c}{(mmag)} &
\multicolumn{1}{c}{(mmag)} &\multicolumn{1}{c}{(mmag)} &  \multicolumn{1}{c}{} & \multicolumn{1}{c}{}\\
\hline
  10574 & 03 32 37.92 &-27 46 09.1 &20.16 &20.16 &   1$\pm$  0 &   4$\pm$  0 &   5$\pm$  0 & -11$\pm$  0 &1 &      \nodata \\
   5444 & 03 32 44.98 &-27 47 36.9 &20.47 &20.49 & -14$\pm$  2 &   3$\pm$  2 &   2$\pm$  2 &   9$\pm$  2 &2 &      \nodata \\
    671 & 03 32 41.09 &-27 48 53.0 &20.60 &20.61 &   3$\pm$  1 &  12$\pm$  1 & -12$\pm$  1 &  -3$\pm$  1 &1 &          Two \\
   8870 & 03 32 36.67 &-27 46 31.1 &20.94 &20.97 & -20$\pm$  3 &  -3$\pm$  3 &  10$\pm$  3 &  13$\pm$  4 &2 &     Two, CXO \\
  10812 & 03 32 39.09 &-27 46 01.8 &21.12 &21.12 &  12$\pm$  1 & -42$\pm$  1 &   4$\pm$  1 &  25$\pm$  1 &4 & Two, CXO, PS \\
      6 & 03 32 39.54 &-27 49 28.4 &21.47 &21.48 &  13$\pm$  1 &  11$\pm$  1 &  -8$\pm$  2 & -17$\pm$  2 &1 &          Two \\
  10330 & 03 32 37.19 &-27 46 08.1 &21.48 &21.51 & -11$\pm$  3 & -19$\pm$  3 &  22$\pm$  3 &   7$\pm$  3 &2 &          Two \\
   9257 & 03 32 44.28 &-27 46 42.3 &22.61 &22.63 &  -7$\pm$ 10 &  14$\pm$ 10 &  20$\pm$ 10 & -28$\pm$ 12 &1 &      \nodata \\
   9719 & 03 32 36.43 &-27 46 32.6 &22.77 &22.82 &   3$\pm$  7 & -43$\pm$  8 &  32$\pm$  7 &   7$\pm$  9 &1 &      \nodata \\
   6489 & 03 32 44.78 &-27 47 24.8 &22.81 &22.81 &  25$\pm$  5 &  10$\pm$  6 & -35$\pm$  6 &  -1$\pm$  7 &1 &      \nodata \\
   1837 & 03 32 33.12 &-27 48 29.6 &23.08 &23.12 & -49$\pm$  6 &  29$\pm$  6 &  -7$\pm$  6 &  25$\pm$  7 &2 &      \nodata \\
   7474 & 03 32 31.51 &-27 47 12.3 &24.21 &24.37 &  97$\pm$ 17 &  87$\pm$ 19 & -55$\pm$ 20 &-147$\pm$ 27 &4 &          Two \\
   8104 & 03 32 42.86 &-27 47 02.7 &24.51 &24.53 & -62$\pm$ 15 & -60$\pm$ 16 &  95$\pm$ 13 &  19$\pm$ 17 &2 &         CXO? \\
   8475 & 03 32 38.99 &-27 46 56.7 &24.76 &24.95 &  11$\pm$ 24 & -57$\pm$ 28 & 133$\pm$ 21 &-101$\pm$ 32 &2 &      \nodata \\
   8145 & 03 32 42.83 &-27 47 02.5 &24.79 &24.78 & -81$\pm$  6 & -88$\pm$  7 &  89$\pm$  6 &  67$\pm$  7 &2 &Two, CXO?, PS \\
   7179 & 03 32 30.17 &-27 47 16.9 &24.80 &24.87 &  64$\pm$ 26 &  50$\pm$ 29 &-136$\pm$ 34 &  11$\pm$ 33 &2 &      \nodata \\
   5802 & 03 32 37.49 &-27 47 31.4 &24.86 &25.46 &   2$\pm$ 49 &  31$\pm$ 51 & 132$\pm$ 44 &-188$\pm$ 72 &1 &      \nodata \\
   6603 & 03 32 48.21 &-27 47 24.1 &25.02 &25.20 &  75$\pm$ 30 &  40$\pm$ 34 &-140$\pm$ 37 &  13$\pm$ 39 &1 &      \nodata \\
   8197 & 03 32 31.83 &-27 47 02.9 &25.11 &25.17 & -86$\pm$ 37 &  19$\pm$ 37 & -60$\pm$ 37 & 116$\pm$ 38 &1 &      \nodata \\
   5299 & 03 32 36.05 &-27 47 37.8 &25.16 &25.18 & -56$\pm$ 35 & 147$\pm$ 32 &   1$\pm$ 34 &-109$\pm$ 45 &2 &      \nodata \\
   6378 & 03 32 36.20 &-27 47 26.2 &25.43 &25.47 & -20$\pm$ 42 & 119$\pm$ 40 &  74$\pm$ 40 &-200$\pm$ 62 &2 &      \nodata \\
   5576 & 03 32 42.23 &-27 47 33.4 &25.55 &25.56 &  50$\pm$ 37 & -14$\pm$ 42 &  87$\pm$ 36 &-135$\pm$ 53 &1 &      \nodata \\
   9306 & 03 32 41.09 &-27 46 42.4 &25.78 &25.77 & 153$\pm$ 33 &-252$\pm$ 53 &  99$\pm$ 35 & -43$\pm$ 49 &2 &      \nodata \\
   4352 & 03 32 33.93 &-27 47 49.9 &25.89 &25.94 & 119$\pm$ 41 &-172$\pm$ 59 & -50$\pm$ 49 &  78$\pm$ 52 &1 &      \nodata \\
   6203 & 03 32 37.00 &-27 47 26.3 &25.92 &26.59 & 228$\pm$ 73 &  60$\pm$ 92 &-482$\pm$141 &  71$\pm$103 &3 &      \nodata \\
   4094 & 03 32 33.11 &-27 47 52.1 &25.96 &25.95 & -50$\pm$ 34 &-247$\pm$ 44 &  84$\pm$ 31 & 169$\pm$ 34 &1 &          Two \\
   6480 & 03 32 41.40 &-27 47 23.9 &25.97 &26.08 &-241$\pm$ 69 & -49$\pm$ 63 & 104$\pm$ 51 & 145$\pm$ 59 &1 &      \nodata \\
   5692 & 03 32 44.97 &-27 47 33.6 &26.03 &26.08 & 191$\pm$ 49 &  82$\pm$ 59 &-179$\pm$ 71 &-137$\pm$ 81 &1 &          Two \\
   4736 & 03 32 43.25 &-27 47 44.0 &26.15 &26.27 &  36$\pm$ 42 &  81$\pm$ 44 & 161$\pm$ 39 &-343$\pm$ 73 &2 &      \nodata \\
   1770 & 03 32 35.75 &-27 48 31.5 &26.32 &26.30 & -24$\pm$ 54 & 173$\pm$ 50 &-274$\pm$ 72 &  76$\pm$ 63 &2 &      \nodata \\
   8370 & 03 32 41.97 &-27 46 58.1 &26.36 &26.48 & -59$\pm$ 78 & 186$\pm$ 67 & 127$\pm$ 66 &-323$\pm$121 &2 &      \nodata \\
   5652 & 03 32 42.05 &-27 47 33.1 &26.59 &26.61 &-342$\pm$ 79 &  46$\pm$ 60 &-105$\pm$ 65 & 301$\pm$ 54 &2 &      \nodata \\
   7882 & 03 32 45.76 &-27 47 05.7 &26.73 &26.67 & 115$\pm$ 87 &-322$\pm$142 & 238$\pm$ 79 &-113$\pm$131 &2 &      \nodata \\
   2511 & 03 32 44.36 &-27 48 16.2 &26.76 &26.90 &-283$\pm$124 &-213$\pm$126 &  66$\pm$ 93 & 321$\pm$ 89 &2 &          Two \\
   9286 & 03 32 45.23 &-27 46 42.5 &26.90 &26.95 & -60$\pm$ 94 & -22$\pm$ 98 & 306$\pm$ 68 &-316$\pm$145 &1 &      \nodata \\
   3990 & 03 32 33.14 &-27 47 52.7 &26.91 &26.89 & -41$\pm$ 64 &-411$\pm$ 97 & 127$\pm$ 56 & 222$\pm$ 61 &1 &      \nodata \\
   6251 & 03 32 41.05 &-27 47 25.7 &26.98 &27.03 &-190$\pm$114 &-408$\pm$149 & 217$\pm$ 78 & 244$\pm$ 91 &2 &          Two \\
   3670 & 03 32 35.94 &-27 47 58.0 &27.04 &27.52 & 199$\pm$139 &-575$\pm$309 & 403$\pm$117 &-299$\pm$268 &3 &      \nodata \\
   4680 & 03 32 38.62 &-27 47 44.7 &27.16 &27.24 & -49$\pm$104 &  20$\pm$105 &-331$\pm$139 & 275$\pm$ 96 &1 &      \nodata \\
   7964 & 03 32 44.27 &-27 47 03.6 &27.20 &27.25 &-386$\pm$156 &   9$\pm$118 & 346$\pm$ 82 & -96$\pm$147 &1 &      \nodata \\
  11336 & 03 32 39.68 &-27 45 46.0 &27.22 &27.25 & -38$\pm$104 & 344$\pm$ 80 &-310$\pm$133 &-101$\pm$132 &1 &      \nodata \\
   5936 & 03 32 45.20 &-27 47 29.1 &27.49 &27.61 &-543$\pm$219 &-127$\pm$162 & 213$\pm$113 & 274$\pm$128 &1 &      \nodata \\
   7094 & 03 32 44.26 &-27 47 14.9 &27.59 &27.58 & 404$\pm$ 78 & 104$\pm$112 &-299$\pm$153 &-404$\pm$203 &1 &          Two \\
   2243 & 03 32 35.93 &-27 48 20.5 &27.69 &27.69 &-110$\pm$168 & 432$\pm$110 & 133$\pm$140 &-802$\pm$396 &2 &      \nodata \\
   1136 & 03 32 37.58 &-27 48 46.3 &27.82 &27.90 & 428$\pm$122 & 142$\pm$171 &-119$\pm$211 &-793$\pm$476 &1 &      \nodata \\
\hline
\end{tabular*}
\tablenotetext{a}{Measured magnitude from total UDF stack.}
\tablenotetext{b}{Magnitude of average flux over each of the four epochs.}
\tablenotetext{c}{Difference between average magnitude and magnitude from Epoch $N$ in milli-mags.}
\tablenotetext{d}{Number of epoch pairs where variability detected (max is 6).}
\tablenotetext{e}{Two--also detected in two epoch test; CXO--located at position of CXO source; PS--point source (see \S~\ref{otherwave}).}
\end{table}

\end{document}